\documentclass{iaus}
\usepackage{graphicx}

\title[ALFALFA in the Leo Region] 
{ALFALFA in the Leo Region: Looking for Missing Satellites in HI}

\author[Stierwalt]   
{Sabrina Stierwalt$^1$}

\affiliation{$^1$ Center for Radiophysics and Space Research,
Cornell University, Ithaca, NY 14853, \break email: sabrina@astro.cornell.edu}

\pubyear{2007}
\volume{244}  
\pagerange{xxx-xxx}
\date{?? and in revised form ??}
\jname{Proceedings Dark Galaxies $\&$ Lost Baryons IAU Symposium}
\editors{J.I. Davies and M.D. Disney, eds.}

\def\etal{{\it et al.}}

\def\msun{$M_\odot$}

\begin{document}

\maketitle

\begin{abstract}
The location of two nearby galaxy groups within $\sim$20 Mpc in the Leo region allows for a detailed study of low-mass galaxies.  A catalog of HI line detections in Leo ($9h36m<\alpha<11h36m, +8^{\circ}<\delta<+16^{\circ}$) has been made from the blind HI survey ALFALFA.  More sensitive single-pixel Arecibo observations targeted Leo dwarf candidates noted optically by Karachentsev et al 2004
(K04) to determine group members and allow for a comparison of HI and optically-selected samples.  This presentation
highlights the differences between the two samples and the significant contribution blind HI surveys can make to the missing satellites problem.

\keywords{
galaxies: distances and redshifts,
galaxies: dwarf,
galaxies: evolution,
galaxies: formation,
galaxies: mass function,
galaxies: spiral,
(cosmology:) cosmological parameters,
cosmology: observations,
(cosmology:) large-scale structure of universe,
radio lines: galaxies}
\end{abstract}

\firstsection 
\section{The ALFALFA Survey of the Leo Region}\label{sec:alf}
As described by Giovanelli, Haynes and others in these proceedings, the blind HI survey Arecibo Legacy Fast ALFA (ALFALFA) utilizes the Arecibo Observatory's
new 7-beam feed array to map 7000 deg$^{2}$ of sky out to cz$\sim$18,000 kms$^{-1}$.  One of the first areas to be covered in full by
ALFALFA was the Leo region, and in these 240 deg$^{2}$ of sky ALFALFA has produced roughly 1300 good quality HI detections, 75$\%$ of which are new HI detections. 
Roughly 40 are found to have HI masses $<$ 10$^{7.5}$\msun.  Complex structure in velocity space at cz $<$3000 kms$^{-1}$ forms 2 distinct entities, the Leo I and Leo II groups.  The nearer Leo I group (also the M96 group) is placed at 10.4 Mpc using Cepheid variables and surface brightness fluctuations.  Leo I offers a uniquely intermediate density environment with both a low velocity dispersion ($\sim$130 km/s) and a local density enhancement high enough to support E/S0 galaxies.  Within Leo I exists further substructure including a ring of intergalactic gas $\sim$200 kpc in diameter surrounding the M96 group (Schneider et al 1983), and a long tidal stream of stars and gas produced by a galaxy interaction in the Leo Triplet.  Leo I also differs from most nearby groups in its comparatively larger distance of 3-4 Mpc from the supergalactic plane (K04).

\section{HI and Optical Comparisons}\label{sec:comp}
A smaller region from $10h30m$ to $11h5m$ in RA was surveyed by K04, and potential Leo
members were selected via a visual inspection of POSS-II/ESO/Serc plates.  To verify group members and create an optically-selected catalog of Leo dwarfs, single-pixel Arecibo observations were taken of the identified candidates.  Twenty of a possible 35 dwarfs were detected in the high-sensitivity HI observations, including 5 background sources, with integration times from 2 to 16 min on-source (compared to ALFALFA's $\sim$40sec/beam). 
Two detections lie in the direction of the Leo Ring and thus
require further analysis to determine the relationship of the stellar dwarf to the detected HI.  

Only three of the optically-selected objects detected in the targeted HI observations were not detected
in ALFALFA because their fluxes lie below ALFALFA's detection limit.  An additional $\sim$20 probable Leo dwarfs were newly identified in the HI catalog.  Four examples of these low-mass objects
are shown in Figure 1.  When placed at the group distance of 10.4 Mpc, AGC202248 and AGC202035 have HI masses less than 10$^8M_{\odot}$ ($\sim 10^{7.21}M_{\odot}$ and
$10^{7.63}M_{\odot}$ respectively), while AGC202024 and AGC205278 are even dwarfier with
HI masses less than 10$^7$\msun ($\sim 10^{6.71}$\msun and $10^{6.94}$\msun respectively).   
\begin{figure}[h]
\centering
\includegraphics[angle=0,width=4in,clip]{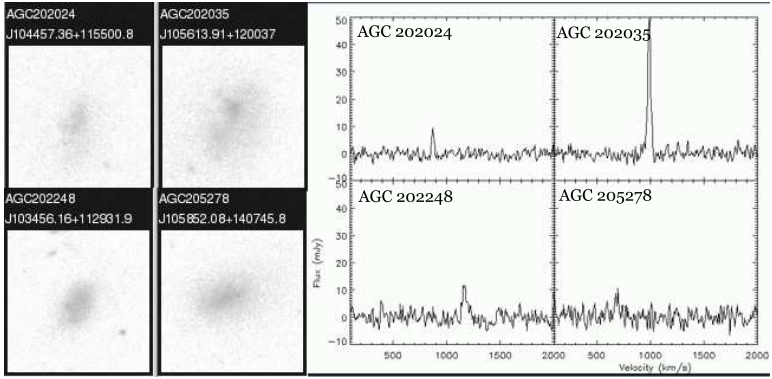}
\label{onlyfig}
\caption{SDSS images (50$^{\prime\prime}$ scale) of 4 Leo dwarfs with their ALFALFA detections.  AGC202024 $\&$ AGC202035
were noted in K04's optically-selected catalog while AGC202248 $\&$ AGC205278 were not.  All have HI masses $<
10^{8}$M$_{\odot}$.}
\end{figure}
\firstsection
\section{Conclusions and Future Work}\label{sec:concl}
The ALFALFA survey has been very effective in finding gas-rich dwarfs that have previously been overlooked by optical
surveys.  With the completion of the ALFALFA Leo catalog, the origins of these dwarfs can now be investigated.  The
gas-rich dwarfs most commonly found in ALFALFA are often optically blue $\&$ clumpy, and many of these dwarfs are very low surface brightness making star formation activity difficult to
determine optically. In such cases sites of star
formation can often still be identified by the faint blue light the youngest stars emit in the ultraviolet.  GALEX
observations of the lowest-mass Leo dwarfs
are currently underway, and the
next stage of this project will be to compare sites of recent star formation 
with HI column densities.

\begin{acknowledgments}
This work has been supported by NSF grants AST--0307661,
AST--0435697 and AST--0607007 and by the Brinson Foundation.  Arecibo Observatory is part of the NAIC which is operated by Cornell University under a cooperative agreement with the NSF.  Funding for SDSS has been provided by the Alfred P. Sloan Foundation, the Participating Institutions, NSF, US DoE, NASA, the Japanese Monbukagakusho, the Max Planck Society, and the Higher Education Funding Council for England.  The SDSS Web site is http://www.sdss.org/.
\end{acknowledgments}

\end{document}